\documentstyle[12pt,aasms4] {article}

\begin{document}

\title{\bf THE EXPECTED DURATION OF GAMMA-RAY BURSTS IN THE IMPULSIVE
HYDRODYNAMIC MODELS}

\author{\it Abhas Mitra\altaffilmark{1}}

\altaffiltext{1}{Theoretical Physics Division, Bhabha Atomic Research Center,
Mumbai- 400 085, India
E-mail: nrl@magnum.barc.ernet.in}

\begin{abstract}

Depending upon the various models and assumptions, the existing literature
on Gamma Ray Bursts (GRBs) mentions that the gross theoretical value of the
duration of the burst in the hydrodynamical models is $ \tau \sim r_{\gamma}/ 2 \eta^2 c$, where
$r_{\gamma}$ is the radius at which the blastwave associated with the
fireball (FB) becomes radiative and sufficiently strong. Here $\eta \equiv
E/Mc^2$, $c$ is the speed of light,  $E$ is initial lab frame energy of
the FB, and $M$ is the baryonic mass of the same (Rees \& Meszaros 1992).
However, within the same basic framework, some authors (like Katz and
Piran) have given $\tau
\sim r_{\gamma}/2 \eta c $.  We intend to remove this confusion by
considering this problem at a level deeper than what has been considered
so far. Our analysis shows that none of the previously quoted expressions
are exactly correct and  in case the FB is produced impulsively and the
radiative processes responsible for the generation of the GRB are
sufficiently fast, its expected duration would be $\tau \sim a r_\gamma/2 \eta^2
c $, where $ a \sim O (10^1) $. We further
discuss the probable change, if any, of this expression, in case the  FB
propagates in an anisotropic fashion. We also discuss
some associated points in the context of the Meszaros \& Rees
scenario.

\end{abstract}


\section{INTRODUCTION}

Our present understanding of the phenomenon of GRBs is based on the
foundations laid by \markcite{CR78} Cavallo \& Rees (1978), \markcite{G86}
Goodman (1986), \markcite{P86} Paczynski (1986), \markcite{E89}
 Eichler et al. (1989),
 \markcite{SP90} Shemi
\& Piran (1990) and several other works. However, as far as the origin of
actually observed nonthermal and highly complex spectra are concerned, we
are indebted to another important idea \markcite{RM92} (Rees \& Meszaros
1992,  \markcite{MR93} Meszaros \& Rees 1993)  that cosmic FBs with
appreciable pollution of baryonic mass, i.e., with a value of $
\eta \sim 10^2 - 10^4 $, where $\eta$ has been defined in the abstract, could be a virtue rather than being a problem.
Most of our current efforts to understand the phenomenology of GRBs
largely hinge on this above mentioned framework.  Meszaros \& Rees (1993)
suggested that the duration of the burst should be $ \tau \sim r_{\rm d}/2
\eta^2 c $, where $r_{\rm d}$ is the so-called deceleration radius measured in
the lab frame. At $r=r_{\rm d}$, the  FB is supposed to transmit
half of its original  momentum to the medium. The baryon polluted FB is expected to drive a strong
forward shock (or a blast wave) in the ambient medium, presumably the
interstellar medium (ISM), and at $ r = r_{\rm d} $, the blast wave is assumed
to be sufficiently strong as the FB transfers half of its momentum to
it.  Technically, we can define another distance, namely, $r_\gamma$,
where either the blast wave or the reverse shock becomes sufficiently
radiative as far as hard X-ray and gamma ray (i.e., the main component of
the observed GRBs) productions are concerned. The
value of $r_\gamma$ will be highly model dependent and contains all the
microphysics of the process, and strictly speaking, there may not be any
simple correlation between the values of $r_{\rm d}$ (largely a simpler
hydrodynamic definition) and $r_\gamma$ (a highly model dependent
definition). We want to emphasize here that the basic definition of $\tau$
should naturally involve $r_\gamma$ and not $r_{\rm d}$.
Unfortunately, the present status of the studies on GRB is
quite preliminary, and, it is not possible to unambiguosly define the
value of $r_\gamma$ even for a relatively simple model. In this situation,
practically, all the authors tacitly assume that $r_\gamma \sim r_{\rm
d}$. Since the present paper endeavours to analyze the question of the
gross time scale in the context of the
existing framework of hydrodynamical model of GRBs we will  also use
the condition $r_\gamma \sim r_{\rm d}$ although we will try to retain the
physical distinction between $r_\gamma$ and $r_{\rm d}$ as far as possible.

 However, following the same basic framework,
\markcite{K94} Katz (1994, see his eqn. 23) finds that for a FB propagating in
a dense ambient medium (which could be a molecular cloud), we should have
$ \tau \sim r_{\rm d}/ \eta c $.  Similarly, \markcite{P94} Piran (1994) also
concludes that $ \tau \sim r_{\rm d}/ \eta c $ (see his eqn. 25).  Although, in a
subsequent paper \markcite{SP95} (Sari \& Piran 1995), the question of
hydrodynamical timescales and temporal structure of GRBs has been
discussed with considerable detail, we feel that the aspect of the gross
overall duration has not been answered in an umambiguous manner. This
foregoing work, in particular, laid specific emphasis on the temporal
structure of the GRBs associated with the occurrence of the Newtonian,
or, subsequently, relativistic, reverse shock  which is supposed to
propagate inside the FB. On the other hand, we would discuss  that, if the
fireball (FB) actually becomes radiative at the deceleration radius (where
it loses half of its initial momentum and kinetic energy to the ambient
medium), the likely evolution of the reverse shock from a Newtonian one to
a fully relativistic one (at $r \gg r_{\rm d}$) could be of somewhat academic
interest.  However, in case the blastwave fails to become radiative at
$r\sim r_{\rm d}$ for some reason or another, the considerations due to
Sari \&
Piran (1995) may become applicable to the actual cases. Here again, one
has to address the question of the overall gross duration of the bursts.

  To fully appreciate the origin of this dilemma and its resolution, we
would revisit the concept of duration of pulses emitted by a FB
propagating in a vacuum.  For the sake of clarity, we would use the
following nomenclature: $\gamma \rightarrow$ the instantaneous bulk
Lorentz factor (LF) of  {\em any} fluid emitting the radiation and
$\gamma_{\rm F}
\rightarrow$ the instantaneous bulk LF of the FB. Thus $\gamma$ is a
general nomenclature applicable to both the original FB (in a vacuum)
 or
any section of the shocked fluid (in a medium) which might be emitting
radiation whereas $\gamma_{\rm F}$ specifically refers to the FB.

Note that, the shock can not be sufficiently strong unless $r >r_{\rm d}$.
We would see that,
in case the GRB eventually results from a shock which becomes
 sufficiently strong as well as radiative at $r =r_{\rm d} = r_{\gamma}$,
as envisaged by Meszaros
\& Rees, we should have $ \tau \sim a r_{\rm d}/2 \eta^2 c $, where $ a
\sim 10 $,
which means that in the Meszaros \& Rees scheme, the actual duration of
the bursts should be at least one order higher than what has been
contemplated so far (for a fixed $\eta$).  We would also try to point out
that as long as the blast wave can be considered sufficiently radiative,
for estimating the eventual duration of the burst,
the reverse shock plays an insignificant role.

\section{FIREBALL IN A VACUUM}

 When a FB propagates in a vacuum, we have $ \gamma
\equiv \gamma_{\rm F}$.  It is interesting to compare the radiation emitted
by the spherically expanding FB with that of a relativistically
moving point source showing apparent superluminal (transverse)
motion   \markcite{R66} (Rees 1966):

\begin{equation}
v_{\perp} = {v \sin \theta \over {1- \beta \cos \theta}}
\end{equation}

 where $ \beta = v/c $, $ \gamma = (1 - \beta^2)^{-1/2} $,
and $\theta$ is the angle between the line of sight and the direction of
motion.  We may obtain $ v_{\perp} > c $ in some circumstances essentially
because the source of radiation moves with a speed $ v \rightarrow c $, and
 tends to catch up with the radiation emitted by
itself.  This phenomenon occurs because the
velocity of propagation of the radiation from the fast moving source
remains fixed at $ c$ and does not increase in a Galilean fashion.  The
same relativistic phenomenon is actually responsible for offering a value
of $\tau$ much shorter than the expected non-relativistic value $\sim
r_{\gamma}/c$.  Nevertheless, for considering the (apparent) time seen by
the observer, we have to consider the line of sight velocity of the FB
rather than the tranverse velocity:

\begin{equation}
v_{\parallel} = {{v \cos \theta} \over {1 - \beta \cos \theta}}
\end{equation}

 For the exact point of the FB intersected by the line of sight, i.e. for
$\theta=0$,  one can see that $ v_{\parallel} = v/(1-\beta)
\approx 2 ~v~ \gamma_{\rm F}^2 $ for $ \gamma_{\rm F} >> 1 $.
Hence $ \tau \sim r/v_{\parallel} \sim r/2 c \gamma_{\rm F}^2 $.  However, the
observer receives light not only from a given point but also from other
parts of the FB.  Because of relativistic aberration, the angular extent
of the region is limited to $ \theta \sim 1/
\gamma_{\rm F}$ in the lab frame.  Then we have

\begin{equation}
\tau \sim {r \over c} {{1- \beta \cos \theta} \over {\cos \theta}} \sim {r
\over {c~ \gamma_{\rm F}^2}}
\end{equation}

 Here it is assumed that $r=r_{\gamma}\gg r_0$,  the initial
radius of the FB. Initially the FB could be non-relativistic and in case it
could manange to send radiation outside (actually it would be extremely
optically thick), the duration of the burst would have been $\tau_0 \sim r_0/c$.
 If the explosive energy is
liberated not instantaneously, but over a tiny but finite
time-scale,  the impulsive approximation is correct as long
as the expected $ \tau \gg \tau_0 $.  If the FB really becomes
optically thin, i.e., the Thomson scattering optical depth of
the FB fluid becomes $ \leq 1$ at $ r = r_{\rm T} $, most of the
radiation escapes out over a distance $ r \sim r_{\rm T} $, and we
would have $ \tau \sim r/c \gamma_{\rm T}^2 $, where $ \gamma_{\rm T} $ is
the bulk Lorentz factor of the FB at this point.  Dynamics of
the FB shows that from the initial non-relativistic phase, the
FB becomes quickly relativistic  and $ \gamma_{\rm F} \propto
 r $, and $ \gamma_{\rm F} \rightarrow
(\eta + 1) \approx \eta $ when $ r_{\eta} = 2~ r_0~ \eta $, and
then the FB coasts freely with $ \gamma_{\rm F} \approx\eta$ (\markcite{PSM93}
Piran, Shemi \& Narayanan 1993, \markcite{MLR 93} Meszaros, Laguna, \&
Rees 1993).  For a wide range of parameters of cosmic FBs, it also follows
that $ r_{\rm T} >> r_{\eta} $, so that $ \gamma_{\rm T} \approx \eta $.
At the same time, the observed duration of the burst cannot be smaller
than $ \tau_0 $.  Therefore, for the sake of consistency, we will have
$\tau = r_{\rm T}/c \eta^2$ if $r_{\rm T}\ge r_0 \eta^2$ and $\tau=r_0/c$
otherwise.

 This exercise suggests that, in the lab frame, the FB
appears as a narrow shell of width $ \Delta r \sim r_{\rm T}/{\gamma_{\rm F}}^2 $ (if
$ r_{\rm T} > r_0 \eta^2 $).  The Doppler factor associated with the
superluminal motion is $ \delta = [\gamma (1 - \beta \cos
\theta)]^{-1} $ and the the comoving duration of the pulse is (see
eq. 2)

\begin{equation}
\tau_c \approx \tau \delta \approx {r\over v}{1 \over {\gamma_{\rm F} \cos
\theta}} \approx {r \over {c \gamma_{\rm F}}}
\end{equation}

 Of course, this result could have been
obtained directly by considering the Lorentz contraction of the length
$ r $ in the fluid frame.  For the baryon polluted FBs with $
\eta < 10^5 E_{51} r_{0,{\rm F}}^{-2/3} $, the escape of radiation at $
r \approx r_{\rm T} $ is only of pedagogic importance because only an
insignificant fraction of the FB energy $ \sim E/ \eta $ is available in the form
of FB radiation at $ r \approx r_{\eta} < r_{\rm T} $.  Almost the
entire energy of the FB gets channelized into the bulk kinetic
energy of the baryons and which cannot radiate efficiently in vacuum
(Shemi \& Piran 1990).

\section{FIREBALL IN A MEDIUM}

 Meszaros \& Rees (1993) pointed out that the baryonic FB
should sweep the ambient ISM and drive a strong forward shock.
In principle, this blast wave may be radiating all the time, but the rate
of radiation cannot be substantial unless appreciable amount of
the FB momentum and energy have been transmitted to the medium and the
shock becomes radiative at $r=r_{\gamma}$. It is also likely that the FB
radiates considerable amount of energy in non-thermal gamma rays even before
$r=r_{\gamma}$ owing to the existence of internal shocks (Rees \& Meszaros
1994). However, we implicitly assume here that the fraction of initial energy
lost in this way would be $<50\%$.

  We can simplify the BF-shock configuration by a one
dimensional sketch following Katz (1994) and Piran (1994) (see Fig. 1).  The region (4)
represents the unperturbed FB whose original edge $S$ is the contact
discontinuity between the piston driving the shock and the
unperturbed ISM (1).  The region (2) is the perturbed and
squeezed-shocked ISM whereas region (3) is the perturbed and
reverse-shocked FB.  Let $ w =e+p $ denote the
proper enthalpy density for each region (with
appropriate subscript) where $e$ denotes the proper  internal
energy density and $p$ represents the pressure.  Again, some clarifications about
the nomenclatures would be in order here. A subscript to the quantities
such as $e$, $p$ and $w$ will represent respective proper values in a
given region, i.e. $e_1$ will represent internal energy density of region
1. In contrast, since, LFs are always meaningful with respect to a certain
inertial frame, $\gamma_{12}$ will represent the value of LF of the region
2 with respect to the region 1 and {\it vice-versa}. Since, now {\em the
radiation is emitted not by the FB but by the shocked fluid in
the region (2) and (3) with a lab frame bulk LF of $
\gamma_{12}=\gamma_{31}=\gamma $} (the lab frame  being the ISM at
rest [1]) {\em it is the value of $
\gamma_{12} $ rather than $ \gamma_{\rm F} $ which now determines the
value of $\tau$}.  And it is here that the value of $\tau$
{\em bifurcates between $ r_{\gamma} / c \eta $ and $ r_{\gamma} / c \eta^2 $ in
the literature}.  Therefore, we must unambiguously find the value
of $ \gamma_{12} $ in terms of $ \eta $ to settle this issue.
Here it may also be emphasized that at least for the 1-D
simplification employed by us the lab frame LF of the
reverse-shocked fluid in region (3) is also $ \gamma_{12} $
though the LF of it with respect to the region (4) is $
\gamma_{34} $ could be very much different from $\gamma_{31}$.

 Now we can recall the strong shock jump conditions
from \markcite{T49} Taub (1949) and \markcite{BM76} Blandford \& Mckee (1976)
and apply the same at $S_1$, the forward shock:

\begin{equation}
{{e_2} \over {n_2}} = \gamma_{12} {{w_1} \over {n_1}},
\end{equation}

\begin{equation}
{n_2 \over n_1} =  {\gamma_{12} \Gamma_2 + 1 \over {\Gamma_2 - 1}},
\end{equation}
and,

\begin{equation}
{\gamma_{S_1}}^2 = {{(\gamma_{12} + 1) [\Gamma_{2} (\gamma_{12} - 1)
+ 1]^2} \over {\Gamma_{2} (2 - \Gamma_2) (\gamma_{12} - 1) + 2}}
\end{equation}
\noindent where $\gamma_{S_1}$ is the LF of the interface $S_1$ between 1 and 2.  Here $ \Gamma$ is defined by the relation
\begin{equation}
p \equiv (\Gamma - 1) (e - \rho)
\end{equation}

 where $\rho$ is the rest mass density in the respective regions
(Blandford \& Mckee 1976).  For a simple one-component fluid, physically
$\Gamma$ is just the ratio of specific heats, and has a value lying
between $4 \over 3$ and $ 5 \over 3 $.  We expect the (forward) shocked
fluid to be highly compressed, heated and to be highly relativistic
(internal energy wise), so that, $\Gamma_2 = 4/3$ and $e_2 = p_2/3$.
On the other hand, the region (1), i.e., the unperturbed ambient medium is
assumed to be cold so that $p_1 \approx 0$ and $ w_1 \approx e_1 \approx m
n_1 c^2 $, where $m$ stands for proton mass.  Then it promptly follows that

\begin{equation}
e_2 \approx m~ n_2 ~c^2~\gamma_{12},
\end{equation}

\begin{equation}
n_2 \approx 4 ~\gamma_{12}~ n_1,
\end{equation}
and,
\begin{equation}
\gamma_{S_1} \approx \sqrt {2}~ \gamma_{12}.
\end{equation}

 If we assume the surface of contact discontinuity to be at perfect
pressure equilibrium then we will have $ p_3 = p_2 $, and, further,
assuming the reverse-shocked fluid  also to be highly relativistic (as far
as internal energy is concerned), i.e., $ p_3 = e_3/3$, we find $e_3
\approx e_2 $.  As to the region (4), i.e., the unshocked part of the FB,
the baryons are assumed to be coasting freely after $ r > r_\eta <<
r_{\gamma} $ with a LF $
\gamma_{\rm F} \rightarrow \eta + 1 \approx \eta $ until they
sacrifice considerable portion (half at $r=r_{\rm d}$) of their bulk energy to
the shocked naterial.  Therefore, at $ r > r_\eta $, the FB material is
also non-relativistic in its rest frame enabling us to write $p_4 \approx
0$, $w_4 \approx e_4 \approx m n_4 c^2$. This allows us to form another
set of simplified jump conditions at $S_2$ :

\begin{equation}
{{e_3} \over {n_3}} \approx \gamma_{34}~ {{w_4} \over {n_4}} = \gamma_{24}~
mc^2
\end{equation}
and
\begin{equation}
{{n_3} \over {n_4}} \approx 4 \gamma_{34} +3 \approx 4 \alpha~ \gamma_{24},
\end{equation}
where
\begin{equation}
\alpha \equiv \left(1 + {3\over \gamma_{34}}\right)
\end{equation}
and the LF of the interface between 3 and 4
\begin{equation}
\gamma_{S_2} \approx \sqrt {2}~ \gamma_{24}.
\end{equation}

The value of $\alpha$ lies between 1 (ultrarelativistic reverse flow, $
\gamma_{34} \gg 1$) and 4 (mildly relativistic  reverse flow,
$\gamma_{34} \approx 1$). Following Piran (1994), if we define $ f \equiv
n_4/n_1 $, and utilize the fact that $\gamma_{\rm F} = \gamma_{12}
\times \gamma_{24} $,  we can eliminate $\gamma_{24} $ from
the foregoing equations to obtain

\begin{equation}
\gamma_{12} \approx f^{1/4}~ \gamma_{\rm F}^{1/2}~ \alpha ^{1/2},
\end{equation}

\begin{equation}
\gamma_{24} = \gamma_{34} \approx f^{-1/4}~ \gamma_{\rm F}^{1/2}~ \alpha^{-1/2},
\end{equation}
and
\begin{equation}
n_3 \approx 4~f^{-1/4}~ {\gamma_{\rm F}}^{1/2}~ \alpha^{1/2}.
\end{equation}

  Katz (1994) has considered a case with $n_4 = n_1$, i.e., $f = 1$, and
$\alpha = 1$ to obtain $ \gamma_{12} \approx {\gamma_{\rm F}}^{1/2}$.  And
thus he obtained $ \tau
\approx r_{\gamma}/c~ \gamma_{12}^2 \approx {{r_{\gamma}}/  {c~
\gamma_{\rm F}}}$.  Let us now try to see whether this consideration of $f=1$ is
justified or not.  As is well understood, each observer sees the FB within
a solid angle of $\gamma_{\rm F}^{-2}$, and a spherical FB will appear as a
collection of $\gamma_{\rm F}^2$ incoherent beamed FBs to as many observers
distributed over $4\pi$ solid angle.  Thus even for a  FB with angular extent of $4 \pi$, we
can actually take care of considerable amount of anisotropic development.
As long as a given ambient medium, which could be the background ISM
$(n_1 \approx 1)$ or a molecular cloud $(n_1 \sim 10^{2-4})$, has a
linear width much larger than the length scales associated with
the development and completion of the radiative processes associated with
a GRB event we can crudely consider the ambient medium to be uniform over
a certain scale.   Since the value of $r_{\gamma}$ is expected to be $<
10^{17}~{\rm cm}$, and the associated solid angle, $\gamma_{\rm F}^{-2}$,
is expected to be $10^{-4}$ to $10^{-8}$ steradian corresponding to
$\gamma_{\rm F} \sim 10^2 - 10^4 $, the linear scales in question are indeed
 much
smaller than the  typical cloud dimensions $\sim$ few pc, and thus,
we should not
be concerned with the probable different values of $n_1$.  Given a certain
fixed value of $n_1$, we can now uniquely find the value of $n_4$ :

\begin{equation}
n_4 = {M \over {4 \pi r^2 (\Delta r)_{\rm com}~ m}}.
\end{equation}

where the comoving width of the FB is $(\Delta r)_{{\rm com}} = r/
\gamma_{\rm F}$. And since $M \approx E/ \eta c^2$, we obtain

\begin{equation}
n_4 = {{E (\gamma_{\rm F}/ \eta)} \over {4 \pi r^3 c^2 m}} \approx
5 \times 10^7~ E_{51}~ r_{15}^{-3}~ (\gamma_{\rm F}/ \eta)~{\rm cm}^{-3}.
\end{equation}

Now, we will have to confront the question: physically, which is the most
appropriate definition for $r_{\gamma}$? If the blast wave becomes
radiative at $r=r_{\rm d}$, obviously, following Meszaros and Rees (1993),
 the radiative radius $r_\gamma$
should be the deceleration radius, $r_{\rm d}$. Simple energy and momentum
conservation consideration shows that at the deceleration radius, i.e., the
radius when half of the initial momentum gets transmitted, we have $
\gamma_{\rm F} =\gamma_{\rm Fd} \approx
\eta/2 $ and the swept up mass $ \approx M/ \gamma_{\rm F} $.  This
yields

\begin{equation}
r_{\rm d} = \left({3 E \over {4 \pi c^2 \gamma_{\rm Fd} \eta m
n_1}}\right)^{1/3} \approx \left({3 E \over {2 \pi c^2 \eta^2
m n_1}}\right)^{1/3} \approx 7
\times 10^{15}  E_{51}^{1/3} \eta_3^{-2/3} n_1^{-2/3} {\rm cm}
\end{equation}
where $\eta_3=\eta/10^3$ and $n_1\rightarrow n_1/1~ {\rm cm}^{-3}$. We can also
contemplate as to which is the fundamental scale length (apart from $r_0$)
in this problem. Could it be the ``Sedov length'' $l\equiv (E/n_1 m c^2)$
where the FB sweeps a mass equal to $E/m c^2$ (Sari \& Piran 1995)? From a
dynamical point of view the concept of momentum exchange is more
meaningful than the concept  of swept up mass and in a
nonrelativistic SNR case, the equality of swept up mass implies equal momentum
sharing.  Therefore in a
relativistic dynamical problem (GRB), it is the ``deceleration length''
which the physical equivalent to the idea of Sedov Length appearing in
the supernova remnant (SNR) theory. As we allow $\gamma_{\rm F} \rightarrow 1$, we find,
$r_{\rm d} \rightarrow l$ within a small numerical factor! If the Sedov Length
were indeed a basic length scale in this problem, there would be a basic
time-scale $l/c \sim 1 ~{\rm yr}$  - a time scale which
is actually appropriate for the SNR case and which may also be appropriate
for low wavelength afterglow following the main GRB.

Of course, it is probable that the shock actually may not be sufficiently
radiative at $r=r_{\rm d}$, but it becomes so at a much later distance
where the value of $f$ would be much lower than $f_{\rm d}$ (in a
spherically symmetric  3-D geometry) and the FB has swept an amount of
mass much larger than $M/{\gamma_{\rm F}}$  because of a variety of
reasons (Sari \& Piran 1995). Unless we have a specific prescription to
describe the extent to which the FB is radiative,  the problem becomes
rather poorly defined in this case. The most important parameter
describing the radiative maturity of the shock could be the {\it in-situ}
magnetic field near the blast wave, and, almost for any model of enhanced
magnetic field generation, the value of the magnetic field decreases at
least linearly until the background  ISM value is achieved (Meszaros,
Rees, and \& Papathanssiou 1994). Therefore, {\em it is highly unlikely
that if the blast wave fails to become radiative at $r=r_{\rm d}$, it
would be radiative at a much larger radius in the framework of a purely
hydrodynamic model}. Nevertheless, we would like point out here that
 this whole discussion explores the question of
duration of the GRB in the idealistic Meszaros \& Rees framework which
assumes that even in a sparse ISM ($n_1 \le 1$), the hydrodynamic limit is
achieved at any radius.  We have
discussed elsewhere (Mitra 1996) that actually this framework may not be be valid at all
when applied to a sparse ambient medium because because
the mean free path of the leading particles of the blast wave ($\lambda$)
is unlikely to satisfy the condition $\lambda \ll \Delta r$, where $\Delta
r \sim r/\gamma^2 c$ is the lab frame width of the FB. This would mean that the
hydrodynamical limit may not at all be achieved at the expected value of
$r\sim r_{\rm d}$ and the FB may not transfer any appreciable amount of
energy and momentum to the ambient medium. Naturally, there may be no strong
shock at all at $r = r_{\rm d}$ (Mitra 1996). In such a case, the FB
may interact with the ambient medium in the fluid limit at a much larger
distance $r\gg r_{\rm d}$  and part of the discussion by Sari \& Piran (1995)
may be applicable in a surprisingly unexpected way.

 For the time being we ignore such disturbing possibility and note that
in any case, it is really
not necessary for the emission of the gamma-rays that the reverse shock
crosses the FB; the blast wave may be radiative enough on its own though
it may look for  seed photons originating from any source including the
reverse shock.  Then at $r = r_{\gamma}= r_{\rm d}$, the foregoing
equations lead to

\begin{equation}
f = {{n_4} \over {n_1}} = {{1} \over {6}} \gamma_{\rm Fd} ~\eta
\approx {{1} \over {3}} \gamma_{\rm Fd}^2 \approx {{1} \over {12}} \eta^2
\end{equation}
 Therefore, we must have $f >> 1$ at the deceleration
radius in the Meszaros and Rees scenario.  Now we can go back
to eqn. (17 - 22) to find that for $r=r_{\rm d}$, we have

\begin{equation}
\gamma_{12} \approx ({1/ 3})^{1/4}~ \alpha^{1/2}~ \gamma_{\rm Fd}
\end{equation}
and
\begin{equation}
\gamma_{24} \approx 3^{1/4}~ \alpha^{-1/2} \approx 1.
\end{equation}

 Because of the uncertainty in the value of the bulk LF
of the reverse shock material (which we clearly find to be $\sim
1$)  we are still not able to exactly fix the value of $\gamma_{12}$.
However, we can do so by appealing to some simple physical
facts.  The first condition is a trivial one  that  we cannot have the
value of any LF, in particular, $\gamma_{24} <1$.
 And in order that there is a forward shock
at all, we must have $\gamma_{S_1} (\approx \sqrt {2} \gamma_{12})
\ge \gamma_{\rm F}$.  Finally, there will be no reverse shock if
$\gamma_{12} \ge \gamma_{\rm F}$.  These physical conditions are
actually so powerful that we could have set the value of
$\gamma_{12}$ to lie between the narrow range of $\gamma_{\rm F}/
\sqrt{2}$ and $\gamma_{\rm F}$ without carrying out much of the
exercise done before.  And the same conditions show that for
$\gamma_{\rm F} \gg 1$, irrespective of the nature of the ambient
medium, we would never have a solution which admits $\gamma_{12}
<< \gamma_{\rm F}$ (for instance $\gamma_{12} \approx \sqrt
{\gamma_{\rm F}}$ discussed before). It is interesting to note that the value of
$\gamma_{12}$ should lie within such a narrow range.
  The physical
constraints also imply that the maximum value of the bulk LF of the
shocked material in the FB is $\sqrt{2}$, and thus the reverse
shock in a GRB problem is bound to be non-relativsitic.  From
such considerations, we can approximately write

\begin{equation}
{{\gamma_{\rm F}} \over \sqrt {2}} \le \gamma_{12} \le \gamma_{\rm F}
\end{equation}
and
\begin{equation}
1 \le \gamma_{24} \le \sqrt {2}
\end{equation}

Now we can go back to the work of Sari \& Piran (1995) to see if the
condition for occurrence of an ultrarelativistic reverse shock is valid or
not. They have considered a case with $f\ll \gamma_{\rm F}^2 $ (see eq. 5
of Sari \& Piran 1995) for which
$\gamma_{34} \gg1$ (see our eq. 16). But, if we use our  eq.(16), we immediately
find that, for such a value of $f$, we would have $\gamma_{12} \ll
\gamma_{\rm F}$,
which is  unphysical in view of the constraint (25). On the other hand,
note that, eq. (22) and (23) are consistent with such physical constraints.

 So, as long as we can assume that the radiative
processes in both the forward-shocked fluid and the reverse-shocked fluid
are sufficiently fast, we have $ \tau \sim r_{\rm d}/c \gamma_{12}^2 $.  But
since the reverse shock is actually, at best, mildly relativistic, particles
are not expected to be accelerated to very high Lorentz factors within the
limited observed duration of GRBs.  And since for the radiative processes
like synchrotron and inverse Compton, the time scales are inversely
proportional to the LF of the particles, there is a possibility that the
reverse-shocked fluid might significantly stretch the expected theoretical
GRB time scales. How  justified is this apprehension?  The strength
of the signal appearing from the two regions (2) and (3) should depend on
the ratio of the  power dissipated in the two regions.  And the
latter should depend on the ratio of the amount of work done by $ S_1 $ on
the ambient medium (1), and by $ S_2 $ on (4).  Because the value of the
fluid pressure is the same at $ S_1 $ and $ S_2 $ ($ p_2 = p_3 = p $), the
rate of compression or the rate of $p dV$ work done by the two shocks is $
\gamma_{S1} : \gamma_{S2}
\approx \gamma_{\rm F} : 1 \approx \eta /2 : 1 $.  Since this rate
is Lorentz invariant, and the expected value of $
\eta_2 \sim 10^2~-~10^4 $, we  find that the amount of
available power that goes into the compression of the FB is
negligible.  Accordingly, the rate of work done by the
reverse-shock in any form, whether it is the heating of the FB,
or producing enhanced magnetic field in it or producing low
energy photons to facilitate inverse Compton boosted gamma-ray
production in the region (2) is actually negligible.

 Thus we come to a very important conclusion that we
can practically ignore the reverse shock in studying the gross time scale
of GRBs within the Meszaros and Rees framework! This above understanding
enables us to write

\begin{equation}
\tau \approx {{a~ r_{\rm d}} \over {2 c \eta^2}}
\end{equation}

\noindent where $ 8 < a <16 $.  This value of $ \tau $ is obviously one
order higher than the usually quoted value of $ \tau \approx
r_{\rm d}/2 c \eta^2 $.  By recalling the value of $ r_{\rm d} $ from eqn.
(21), we can rewrite

\begin{equation}
\tau \approx 0.1 ~a~ E_{51}^{1/3}~ \eta_3^{-8/3} ~n_1^{-1/3} {\rm s}
\end{equation}

\noindent For  a given value of $ \eta $, the above derived
value of $ \tau $ is at least one order large than similar
values used in the literature (for instance see eqn. [5.2] of
Mochkovitch et al, 1995)

\medskip

\section{DISCUSSION}

 We have been able to remove a basic qualitative deficiency in the
theoretical description of GRBs in the Meszaros and Rees scenario, which
has become the ``standard model'' for understanding the phenomenology of
these events, i.e., whether $ \tau \sim r_{\rm d}/2 c
\eta^2 $ or $ \tau \sim r_{\rm d}/2 c \eta $.  Although, none of these
expressions appeared to be quite correct our final result is
obviously tilted in favour of the former relation which is,
however, to be modified by a numerical factor $ 8 < a < 16 $.  We
have also emphasized an important aspect in this problem:
the minimum value of the ratio of the Lorentz invariant power liberated in the forward blast
wave and the reverse shock goes as $ \eta/2 \gg 1 $.  This point
should considerably simplify the evaluation of the emitted
spectrum in the problem.  Although we have assumed a spherical
FB, this discussion remains unchanged even for a case when the
actual FB has a funnel type structure in the proper frame.  To
appreciate this subtle point, consider the creation of jet type
FBs following a  NS-NS  collision event (Rees \& Meszaros
1992). Suppose the $ e^+ + e^- $ FB is  due to
collisions of $ \nu$ and $ \bar{\nu}$s emanating from the
super hot and about-to-merge neutron stars.  Then, despite the
asymmetry in the binary geometry, $ e^+ + e^- $ pairs would be
produced over the full $ 4 \pi $ solid angle, though there would be
excess production along the symmetry axis because of larger
value of collision angle along the same.  Unless we
are sure about the exact configuration of the neutrinosphere
associated with each $NS$, or the configuration of the thick
disk that may be formed as the result of m{\rm erg}er, and also the
temporal evolution of the whole pattern, it can only be a matter
of conjecture as to how much excess production of pairs is
achieved at a particular point with a given value of $r$ and
$\theta$.  The canonical values of $ E \sim 10^{50} $ or $
10^{51} {\rm erg} $ is actually obtained by assuming a broad
isotropic picture of the events, and one should actually
consider the use of a fiducial value of $E$ {\em per unit solid angle} for dealing
the anisotropic cases.  In case there is some excess production in a certain
direction, {\em it does not mean that the pair flux produced in other
directions somehow gets focussed in that  direction}
to result in the same canonical value of $E$ assumed for the
spherical $4 \pi$ case.  And the basic reason that there might be a
funnel type geometry of the FB is not the somewhat excess
production of pair in a given direction.  The basic reason, instead,
 is that the mass flux of the debris of NS-NS merger event
is assumed to be appreciably less in those preferred directions
yielding a value of $\eta \gg 1 $.  If the merger event spews
off a baryonic mass $ \sim 10^{-3} M_\odot $, then even for an
assumed (spherical) value of $ E \sim 10^{51} {\rm erg} $, in the
isotropic case we would have $ \eta \le 1 $, and there would be
no GRB!  It is primarily on this account that the assumption of relatively
baryon free funnel along the symmetry axis gives rise to jet-like FBs.  It
only means that there are other jet-like FBs too but they have
unacceptably low value of $\eta$.  Thus, in our view, once we keep a
canonical value of $E$ and tacitly assume an intrinsically spherical FB
(though having various values of $\eta$ in various directions), for a
given observer, we must not additionally plug in the solid angle factor in
associated quantities like $r_{\rm d}$ or $\tau$, as have been done by Meszaros,
Rees and Papathanassion (1994).  To further appreciate this point, we can
specifically consider two typical equations considered by them (2.4 and
2.16)

\begin{equation}
\tau \sim 1 E_{51}^{1/3}~ \eta_3^{-8/3}~ n_1^{-1/3}
\theta^{-2/3}~ {\rm s}
\end{equation}
and the total burst fluence

\begin{equation}
S_0 = 10^{-6}~ E_{51}~ \theta^{-2} ~D_{28}^{-2}~  {\rm erg}~{\rm cm}^{-2}
\end{equation}

 where $ \theta $ is the semi-angle of the funnel,
$S_0$ is the total bolometric fluence of the burst occuring at a
distance of $D = 10^{28} {\rm cm}$.  Note that, although, both {\em mathematically
and physically $\theta = 0$ should correspond to absence of any
burst, the two foregoing  equations would suggest $\tau
\rightarrow \infty$, and $S_0 \rightarrow \infty$ for a
non-existent  burst!
This happens because the value of
$E$ was not scaled down in keeping with the value of $\theta$ in these
equations virtually presuming that the canonical value for an isotropic
FB energy somehow gets channelized along the narrow funnels}.
Therefore, we reinterpret our basic result that in case the FB
is conical with a solid angle $ \Omega \sim \theta^2 $ in its
rest frame, then $E_{51}$ is to be replaced by $ {{\Omega} \over
{4 \pi}} E_{51}$ to take care of the fractional energy
channelization. It follows then that if we apply equations similar to
(29) or (30), we, eventually get back the original eq.(28). In other words,
the basic temporal properties of the FB (along a given direction) should be more or less unaltered
even for an anisotropic case as long as we are able to define $\eta(\theta)$ in a meaningful way. It may be reminded here that although
the generation of a relatively baryon free, high $\eta$ jet-like FB is
required for understanding the GRBs the understanding about their genesis
is largely a matter of conjecture and is something like having ``an
artist's conception''. This is so in view of the (i) uncertainty about
the physical mechanism triggering GRBs, (ii) even for an assumed mechanism
like a NS-NS merger, the merger geometry is unknown and evolving faster
(at the moment of maximum energy liberation) on timescales shorter than
the observed GRB timescales, (iii) unknown  effects like probable new
general relativistic
instabilities and the unknown nature and evolution of the coelesced object, (iv)
the unknown extreme parameters like temperature, density and their
profiles in the dynamic and unknown merger geometry like accretion disk,
torus etc. etc., (v) the unknown microphysics at such extreme conditions
like the equation of state and viscosity profiles of the dynamic merged object,
and also the basic uncertainty about (vi) the relative importance of
energy loss by gravitational radiation and neutrino emission.

The actual burst duration, even when we consider the emission of {\em hard
X-rays and gamma rays only}, can obviously be longer than what is suggested
by eq. (29) if the FB fails to radiate the available energy ($50 \%$ at
$r=r_{\rm d}$). In fact, when the burst fails to be radiative at $r\sim
r_{\rm d}$, there may be a  weak and prolonged burst corresponding to a
larger value of $r_\gamma$.
Also, if we redefine the burst duration as
the one during which the FB radiates $75\%$ of its energy we would have a
value of $\tau$ larger by a factor of few ($\gamma_{\rm F} \rightarrow
\gamma_{{\rm F}/2}$).

 We remind the reader again that this whole discussion explored the question of
duration of the GRB in the idealistic Meszaros \& Rees framework which
assumes that even in a sparse ISM ($n_1 \le 1$), the hydrodynamic limit is
achieved at any radius.  Since the leading protons of the FB
interact with the sparse ISM
extremely weakly because the bacground ISM magnetic field is very
weak,  and we are not aware of any cooperative phenomemon by
which these protons may collectively interact much more strongly by self
generating strong magnetic fields on their way up, {\em it is quite uncertain}
whether GRBs can be triggered in ordinary sparse ISM
(Mitra 1996).
 Naturally, there may be no strong
shock at all at $r = r_{\rm d}$ unless GRBs are hatched in special dense
regions
of ISM although subsequently the blastwave may propagate in the ordinary
ISM and generate various low energy afterglow.   The eventual
mechanism of GRBs could be considerably different than what our present
understanding admits and then part of the present discussion may be
invalidated. For instance all the existing hydrodynamical models assume
that {\em protons/ions accelerated in the shock can transfer their energy to
the associated electrons sufficiently fast so that the radiative time
scale of the shock is determined by  the much shorter radiative time scale
of the electrons}. This assumption  {\em need not be correct} and though it
was pointed out privately by the present author to several other authors working on
this problem this aspect has been glossed over for the sake of simplicity.
 We will discuss in a seperate paper under what condition the
hydrodynamical description may be valid even at small values of $r$ and
how it may be possible to have a blast wave with a higher efficiency for
gamma ray production (Mitra 1997).

It may be also possible that the fundamental mechamism for the GRBs is not
an one shot collision process (involving compact objects) but, on the
other hand, a process whereby  energy is released erratically and in a
jerky manner in the form of an ``unsteady wind'' over a duration of tens of seconds or even longer. In such a
case the basic GRB duration will obviously be determined by the central source
though afterglow time scales may be determined by hydrodynamic model
discussed in this work.

Finally, let us point out that one can also seperately define $r_{\rm
optical}$, $r_{\rm infrared}$, and $r_{\rm radio}$ in the context of the
hydrodynamical model for the post GRB phase of the event
provided one has a model for generation of respective radiations. After the
main GRB phase, the value of $\gamma$ would decrease rapidly and would
approach unity, i.e, the blast wave will become Newtonian like a SNR. At
each stage the afterglow will be characterized by a time scale $\tau_{\rm
radiation} \sim {r_{\rm radiation}\over \gamma^2 c}$ with no simple and
general relationship between $\tau_{\rm radiation}$ and $\eta$. Such
afterglow time scales can obviously be arbitrarily long and exceed even
hundreds of years if the GRB event occurs in the Local Group. Thus, if we
are fortuitous, we may be able to identify some of the presenly known SNRs
which have no signature of harbouring a compact object at their centers as
Gamma Ray Burst Remnants (GRBR).

\noindent Acknowledgement : The author is thankful to the anonymous referee
for several constructive suggestions which greatly improved the
presentation of this work.

\newpage

\centerline{Figure Caption}

Figure 1: Sketch of the FB-shock configurations, for details, see text.

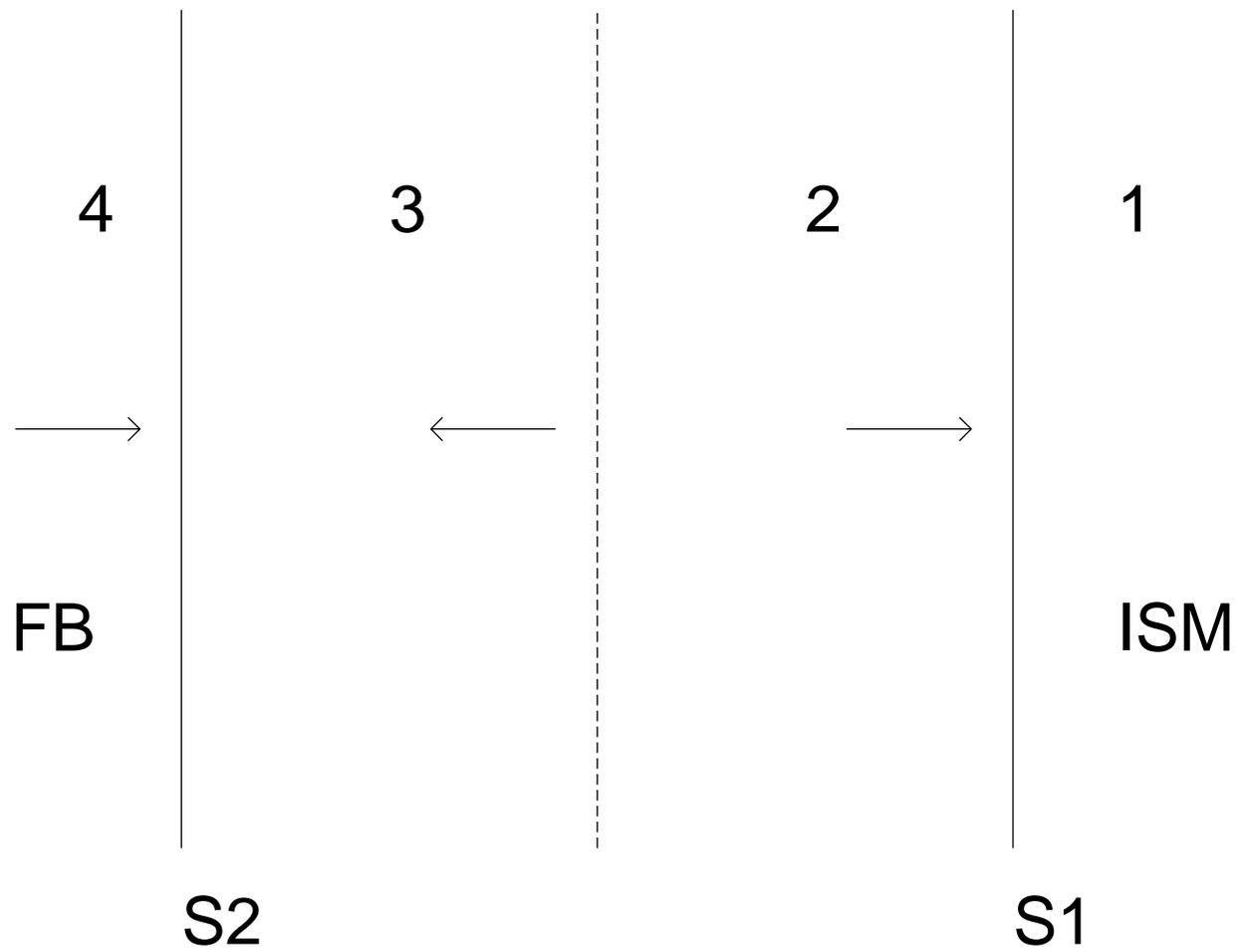
\end{document}